%% file: comrade.tex
\documentclass[conference]{IEEEtran}
\IEEEoverridecommandlockouts
\usepackage{cite}
\usepackage{amsmath,amssymb,amsfonts}
\usepackage{algorithmic}
\usepackage{graphicx}
\usepackage{textcomp}
\usepackage[usenames,dvipsnames,svgnames,table,pdftex,dvipsnamese]{xcolor}
\def\BibTeX{{\rm B\kern-.05em{\sc i\kern-.025em b}\kern-.08em
    T\kern-.1667em\lower.7ex\hbox{E}\kern-.125emX}}

\usepackage{caption}
\usepackage{subcaption}
\usepackage[hidelinks]{hyperref}
\usepackage{siunitx}
\usepackage{makecell}
\usepackage{multirow}

\usepackage[acronym]{glossaries}
\input{acronyms}

\newcommand{\com}{Three-Chains}
\newcommand{\ifunc}{\textit{ifunc}}
\newcommand{\ifuncs}{\textit{ifuncs}}
\newcommand{\am}{Active Message}
\newcommand{\ams}{Active Messages}
\newcommand{\cc}{\textit{Two-Chains}}
\newcommand{\ccc}{\textit{Three-Chains}}
\newcommand{\fatir}{\texttt{fat-bitcode}}
\newcommand{\thor}{Thor}
\newcommand{\ookami}{Ookami}
\newcommand{\jitting}{JIT compilation}
\newcommand{\jitted}{JIT compiled}

\usepackage{dblfloatfix}
\usepackage{tikz}
\newcommand\copyrighttext{
\footnotesize \textcopyright 2022 IEEE. Personal use of this material is permitted. Permission from IEEE must be obtained for all other uses, in any current or future media, including reprinting/republishing this material for advertising or promotional purposes, creating new collective works, for resale or redistribution to servers or lists, or reuse of any copyrighted component of this work in other works.}

\newcommand\copyrightnotice{
\begin{tikzpicture}[remember picture,overlay]
\node[anchor=south,yshift=10pt] at (current page.south) {\fbox{\parbox{\dimexpr\textwidth-\fboxsep-\fboxrule\relax}{\copyrighttext}}};
\end{tikzpicture}
}

\begin{document}
\title{Bring the BitCODE - Moving Compute and Data in Distributed Heterogeneous Systems}

\makeatletter
\newcommand{\linebreakand}{
  \end{@IEEEauthorhalign}
  \hfill\mbox{}\par
  \mbox{}\hfill\begin{@IEEEauthorhalign}
}
\makeatother

\author{
\IEEEauthorblockN{1\textsuperscript{st} Wenbin Lu\textsuperscript{\textsection}}
\IEEEauthorblockA{\textit{Stony Brook University} \\
Stony Brook, NY \\
wenbin.lu@stonybrook.edu}
\and
\IEEEauthorblockN{2\textsuperscript{nd} Luis E. Peña\textsuperscript{\textsection}}
\IEEEauthorblockA{\textit{Arm Research} \\
Austin, TX \\
luis.epena@arm.com}
\and
\IEEEauthorblockN{3\textsuperscript{rd} Pavel Shamis\textsuperscript{\textsection}}
\IEEEauthorblockA{\textit{Arm Research} \\
Austin, TX \\
pavel.shamis@arm.com}
\and
\IEEEauthorblockN{4\textsuperscript{th} Valentin Churavy\textsuperscript{\textsection}}
\IEEEauthorblockA{\textit{MIT} \\
Cambridge, MA \\
vchuravy@mit.edu}
\linebreakand 
\IEEEauthorblockN{5\textsuperscript{th} Barbara Chapman}
\IEEEauthorblockA{\textit{Stony Brook University} \\
Stony Brook, NY \\
barbara.chapman@stonybrook.edu}
\and
\IEEEauthorblockN{6\textsuperscript{th} Steve Poole}
\IEEEauthorblockA{\textit{Los Alamos National Laboratory} \\
Los Alamos, NM \\
swpoole@lanl.gov}
}

\maketitle

\begingroup\renewcommand\thefootnote{\textsection}
\footnotetext{Equal contribution}
\copyrightnotice{}

\input{content/abstract.tex}


\input{content/intro.tex}
\input{content/background.tex}
\input{content/design.tex}
\input{content/experiments.tex}
\input{content/eval.tex}
\input{content/conclusions.tex}
\input{content/ack.tex}

\IEEEtriggeratref{13}
\bibliographystyle{IEEEtran}
\bibliography{IEEEabrv, comrade}

\end{document}

%% file: acronyms.tex
\newacronym{dapc}{DAPC}{Distributed Adaptive Pointer Chasing}
\newacronym{gbpc}{GBPC}{Get-Based Pointer Chasing}
\newacronym{tsi}{TSI}{Target-Side Increment}
\newacronym{xrdma}{X-RDMA}{eXtended Remote Direct Memory Access}
\newacronym{rdma}{RDMA}{Remode Direct Memory Access}
\newacronym{bf2}{BF2}{BlueField-2}

%% file: content/abstract.tex
\begin{abstract}
In this paper, we present a framework for moving compute and data between
processing elements in a distributed heterogeneous system. The implementation of
the framework is based on the LLVM compiler toolchain combined with the UCX
communication framework. The framework can generate binary machine code or LLVM
bitcode for multiple CPU architectures and move the code to remote machines
while dynamically optimizing and linking the code on the target platform. The
remotely injected code can recursively propagate itself to other remote machines
or generate new code.
  
The goal of this paper is threefold: (a) to present an architecture and
implementation of the framework that provides essential infrastructure to
program a new class of disaggregated systems wherein heterogeneous programming
elements such as compute nodes and data processing units (DPUs) are distributed
across the system, (b) to demonstrate how the framework can be integrated with
modern, high-level programming languages such as Julia, and (c) to demonstrate
and evaluate a new class of \acrfull{xrdma} communication operations that are
enabled by this framework. 
  
To evaluate the capabilities of the framework, we used a cluster with Fujitsu
CPUs and heterogeneous cluster with Intel CPUs and \acrlong{bf2} DPUs
interconnected using high-performance \acrshort{rdma} fabric. We demonstrated an
\acrshort{xrdma} pointer chase application that outperforms an \acrshort{rdma}
GET-based implementation by 70\% and is as fast as \ams{}, but does not require
function predeployment on remote platforms. 
\end{abstract}

%% file: content/intro.tex
\section{Introduction}
As Moore's law runs out of steam, hyperscalers are looking for every possible
opportunity to maximize the efficiency of their datacenters. Their pursuit for
efficiency, having now pushed the limits of the SoC, has moved the system
architecture towards disaggregation using low-latency, high-bandwidth
interconnects including PCIe Gen6, CXL\cite{CXL}, CCIX\cite{CCIX}, and converged
and custom Ethernet. Such disaggregation comes at a cost of reduced efficiency
due to data movement overheads and increased security complexity. This reduction
has caused a perfect storm for the emergence of data processing units (DPUs),
computational storage devices (CSDs), and other custom accelerators. The basic
goals of these devices are (a) to move the compute closer to the data in order
to increase efficiency and (b) to enhance security through isolation.

DPU solutions have been adopted by multiple hyperscalers and they are
increasingly seen in datacenter networks. CSDs are the next frontier of
innovation at the disaggregated data center. Despite having the potential for
rich programmable capabilities, both these kinds of devices are primarily used
as fixed-function accelerators with very limited control from the application on
the host. The application cores in these devices are currently limited due to
the software stack not being well established for this domain; the hardware and
software are functional and deployed at scale, but in terms of application
maturity and adoption, both are in their nascent stages and reminiscent of the
early days of GPGPU compute.

Intel has developed the IPDK\cite{IPDK} framework which bundles container with
multiple existing software packages like Open Virtual Switch (OVS) \cite{ovs}
that can be programmed using the P4 programming language\cite{bosshart2014p4}.
Nvidia's DOCA \cite{doca} package also bundles a rich set of software packages
ranging from high-level Snort\cite{snort}, OVS, and NVMe virtualization to
low-level abstraction like rdma-core\cite{rdma-core} and DPDK\cite{DPDK}. In
both cases the software has to be predeployed such that DPDK exposes
fixed-function offload and acceleration. As a result, host based applications do
not have direct access to the compute resources on DPUs.

We believe there are several challenges contributing to the limited exposure of
compute resources to the host applications:
\begin{itemize}
    \item Security constraints -- one of the main use-cases for current DPUs in
    the datacenter is the security-control-point and hypervisor offload for
    bare-metal instances. Therefore, deployment of user application codes
    side-by-side with infrastructure management codes (verified and trusted)
    raise concerns about the impact on system security. In this paper we do not
    focus on addressing security concerns, instead relying on existing
    hardware~\cite{arm-cca} and software confidential compute~\cite{arm-veracruz}\cite{google-oak} efforts.
    \item Lack of standardized interfaces for offloading compute to Arm SoC
    subsystem and embedded accelerators -- this is the main challenge that we
    aim to address with the \ccc{} framework. It is important to note that
    \ccc{} is a path-finding project that aims to demonstrate how such
    interfaces could be implemented and to evaluate potential trade-offs. 
\end{itemize}

In order to democratize programmability of this new class of disaggregated
systems and directly expose programming elements in DPUs and other accelerators
to application layers we developed \ccc{}. The \ccc{} framework is implemented
from scratch, and it is inspired by \cc{} \cite{grodowitz2021cc}. It utilizes
the idea of remote binary code injection while introducing new capabilities and
bringing portability and performance to new levels. \ccc{} provides
infrastructure for moving functions within a heterogeneous computing environment,
caching, remote dynamic linking, and remote recursive invocation. We decided to
name it \ccc{} because it extended original toolchain with LLVM \cite{LLVM}.
\ccc{} leverages the LLVM framework and intermediate representation (IR) for
portable function representation, just-in-time (JIT) compilation, and execution.

Using LLVM IR, the framework implements a multi-architectural function
representation which we call \fatir{}, as well as a separate binary function representation. One of the benefits of using \fatir{} beyond portability is that
the code gets optimized for the target micro-architecture. Since LLVM is used by
multiple programming languages as a backend, \ccc{} thus also supports multiple
programming languages. For this paper we focused on the C and Julia programming
languages \cite{julia} as representative languages for both low- and high-level
programming environments.

Leveraging LLVM, \ccc{} has full support for distributed linking wherein
functions get linked on the target machine. As a result, \ccc{} functions can
interact with external libraries including UCX itself. This also means that
\ccc{}'s remotely injected functions can recursively inject functions to other
processing elements in the system. Such a function can propagate itself to a
remote machine and potentially dynamically select new functions for further
remote injections. Using functions in this way, we have implemented a novel type
of one-sided network remote operations hereafter called \acrfull{xrdma}, in
which \acrshort{rdma} operations trigger recursively executed code inside the
network with processing elements. In order to minimize the overhead associated
with moving code over the fabric, we have also implemented a caching mechanism
that substantially reduces the size of the network messages and avoids the
\jitting{} already-seen code.

The key contributions of the paper are:
\begin{itemize}
    \item High-performance framework for moving and executing complex functions
    using either binary representation or LLVM's intermediate representation
    \item Network protocols and caching algorithms for efficient code delivery
    and multi-architecture code representation
    \item Implementation of dynamic linking for remote injected codes using
    LLVM's ORC-JIT API 
    \item Evaluation of overheads for remote function communication, deployment
    (JIT vs binary), caching, and invocation
    \item Extension of an in-process JIT compiled language (Julia) with remote
    function injection, linking, and execution
    \item Introduction of \acrshort{xrdma} operations, and a demonstration in
    the form of pointer-chase algorithms to DPUs and host processing
\end{itemize}

%% file: content/background.tex
\section{Background}

There are multiple projects and programming models that implement high-performance
RPC message-like semantics~\cite{rpc-chen2019scalable}\cite{rpc-kalia2019datacenter}\cite{rpc-li2021hatrpc}.
The \com{} project was developed from scratch while leveraging concepts from
earlier \cc{} research \cite{grodowitz2021cc}\cite{pena2021ucx}.
The primary difference between our work and the previous state of the art is
that (a) our implementation moves the code and data while supporting multiple
code representations and (b) it leverages JIT techniques to support optimizations
for heterogeneous systems. 

Below, we highlight some of the projects that aim to solve a similar class of problems.
These projects are GASNet~\cite{bonachea2018gasnet},
Google's Snap microkernel~\cite{snap2019}, Charm++~\cite{acun2014parallel},
CHAMELEON~\cite{Klinkenberg2020Chameleon}, FaRM~\cite{dragojevic2014farm} and
Julia~\cite{Besard2019-zu, rizvi2021julia}. GASNet is a communication library
used by the HPC community as a communication backend for programming languages
like UCP, Chapel, and others. GASNet provides APIs for registering and invoking
active messages, but those require code presence on the remote machines. This
differs from our approach where the code is communicated over the network and
dynamically linked and optimized on the target platform without requiring
compile-time registration. Google's Snap Microkernel project provides a platform for remote
procedure calls in the context of network functionality distribution but it
functions as an OS extension whereas \ccc{} is implemented as a user space
library. Charm++ is a distributed programming model that defines distributed C++
objects with a unified logical view and the ability to call methods on those
objects. It implements a task scheduler, automatic object migrations, a run-time
for task launch, communication channels, among other features. The CHAMELEON
framework  uses compiler \texttt{\#pragma} and runtime APIs to define OpenMP
tasks as entities that can be moved around the distributed system using its
built-in task scheduler. Conceptually, CHAMELEON's approach is similar to
Charm++, and our framework can serve both programming models as a foundational
abstraction for object/task migration.
The FaRM project implements a distributed computing platform for distributed
shared memory programming. It uses \acrshort{rdma} network for remote object
manipulation, but unlike our work it does not provide functionality for moving
compute.
In \ccc{}, the Julia programming language is used to demonstrate how to integrate
with high-level programming models and languages. Julia provides a remote procedure
call (RPC) library, called \texttt{Distributed.jl}, in the standard library.
It allows the user to send unoptimized functions to a remote process,
and the function is then optimized and compiled on the remote process; in contrast,
\ccc{} moves optimized bitcode emitted for the target architecture. Furthermore \texttt{Distributed.jl} does
not take advantage of \acrshort{rdma} networks. LLVM's ORC JIT supports a remote executor protocol that allows
for JIT compilation from one process into the other as well as executing remotely compiled functions; in contrast, \ccc{} provides a
mechanism for propagation of bitcode allowing functions to propagate across a
cluster.

%% file: content/design.tex
\section{Design and Implementation}
The primary problem statement that \ccc{} addresses is considering heterogeneous
distributed system design, like the one that can be found in modern datacenters
with GPUs, DPUs, and other processing elements, how one can deploy and execute
application-defined functions. We believe that for a software framework to
address the above problem, it has to provide the following functionality:
\begin{itemize}
    \item A workflow and APIs for moving code and data.
    \item An infrastructure for a dynamic linking of dependencies on the target platform.
    \item Support for invoking remote or local functions.
    \item Support for multiple CPU architectures.
\end{itemize}
It is important to point out that this is very different from RPC and \ams{}
communication semantics, which explicitly assume that the target procedure call
is present on the remote machine with all its dependencies.

\subsection{\com{} Workflow}
The \ccc{} framework provides a set of C APIs that defines how to create and
move executable code in a distributed system. \ccc{} uses the UCX\cite{shamis2015ucx}
communication framework for moving code and data around the system. An
executable piece of code in \ccc{} is called an \ifunc{}, which stands for
"injected function". In addition to the \ifunc{} code, the user can also send a
contiguous chunk of memory, called payload, from the source process to the
target process. The payload contains data that is needed by the \ifunc{} when it
is executing on the target process. The \ccc{} API is implemented as an
extension of the UCP interface in UCX, and it is inspired by API design
described here \cite{pena2021ucx}. Due to space constraints, we do not include
the API description here and focus on the workflow for application codes.

\begin{figure}[h]\includegraphics[width=0.75\columnwidth]{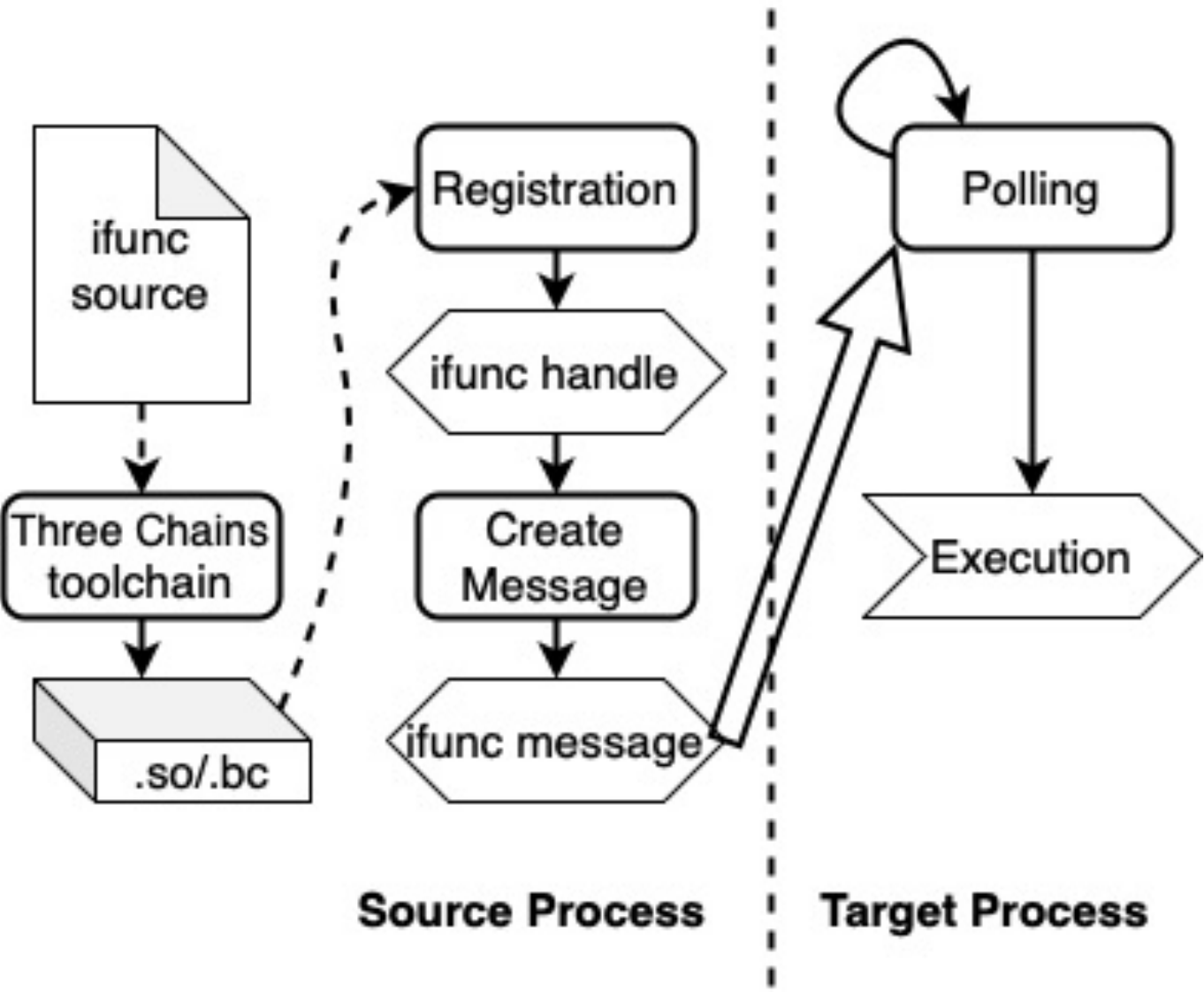}
 \centering
 \caption{\ccc{} workflow}
 \label{fig:workflow}
\end{figure}

A brief description of the \com{} workflow is shown in \autoref{fig:workflow}
and is described as follows. The application developer creates an \ifunc{}
library that implements an entry function (main) and several other functions
required by the \ccc{} framework. Once the \ifunc{} is processed using the
\ccc{} toolchain, the generated files should be placed in a directory that can
be located by \ccc{}. Then, in the application, the \ifunc{} library is
registered using its name (e.g., "foo"), and a handle is returned to the
application. With this handle, the user can create and send \ifunc{} messages of
type "foo" to a target process element.

The target process should use an UCX \ifunc{} polling function to check the
message buffer and handle incoming \ifunc{} messages. Once a message has arrived,
the polling function will load the \ifunc{} code to the target process's address
space and invoke the entry function with a pointer to the payload, and a
user-defined target pointer. Ideally, the target processes should setup a daemon
thread that polls the message buffers periodically. The dynamic nature of
\ifunc{}s allows adding new functionalities to the target process without
recompiling the application and restarting it.

\subsection{The Challenges of the Binary-Based \ifunc{} Implementation}\label{sec:ifunc-binary}

\begin{figure}[h]\includegraphics[width=0.9\columnwidth]{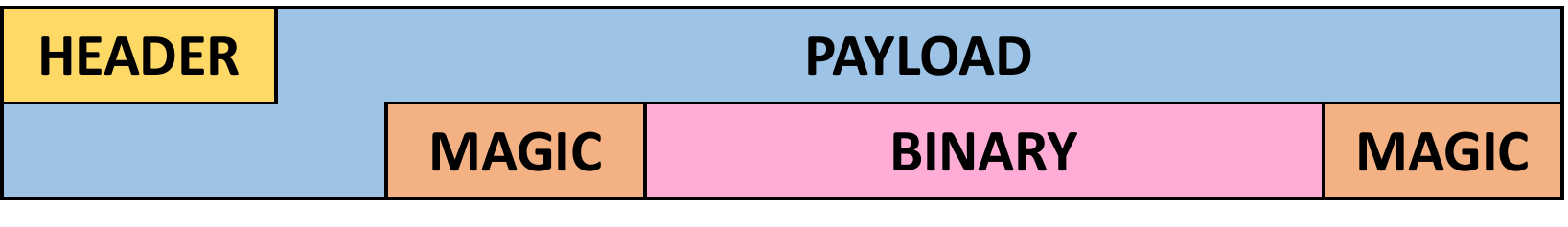}
 \centering
 \caption{Message frame of binary-based \ifuncs{}. HEADER carries information that describes that type and format of the message, MAGIC field is used to discover and identify message delivery, BINARY fields correspond to binary code.}
 \label{fig:ifunc_msg_so}
\end{figure}

The original implementation of the \ifunc{} concept is based on a standard ELF
binary library format. When the user registers an \ifunc{}, the runtime loads
the \texttt{.so} file using \texttt{dlopen()}. To send a message, the
\texttt{.text} and \texttt{.data} sections of the \ifunc{} dynamic library are
copied directly from the dynamic library and packed into a contiguous buffer
along with the payload and some other metadata, as shown in~\autoref{fig:ifunc_msg_so}.

The major challenge of this approach is that run-time symbol resolution must be
performed again on the target process, i.e., one has to do remote dynamic
linking. In our previous work~\cite{pena2021ucx}, the remote dynamic linking is
implemented by redirecting all Global Offset Table (GOT) references to a special
location, which will contain the GOT reconstructed by the target process's
runtime, and so all symbols will resolve to the correct address. The
reconstruction requires the presence of a dynamic library \texttt{.so} file on
the target process's file system, and the use of \texttt{ld.so} linker to
perform relocation of the symbols. If the \ifunc{} does not use external symbols,
then it is defined as "pure”, and we skip GOT patching and go straight to
execution. The details of the binary-based implementation can be found in~\cite{pena2021ucx}.

The above approach has several issues. First, the binary code of the \ifunc{}
dynamic library is instruction set architecture (ISA)-specific, which means the
\texttt{.so} file for a x86\_64 host cannot be used to send \ifunc{} messages to
an Arm-based SmartNIC. This issue could be solved by setting up cross-compilation
toolchain to compile the \ifunc{} for different ISAs/$\mu$Archs and select the
appropriate one when creating an \ifunc{} message. Function-multiversioning
could even pack implementations of the same function specialized for different
$\mu$Archs into the same binary, so we can have \ifunc{} specialized for
different $\mu$Arch of the same ISA, in a single \texttt{.so} file. But overall,
solutions to this problem tend to greatly increase the complexity of both the
\ccc{} toolchain and the user's workflow.

The second issue is that, patching the assembly code for GOT redirections and
remote dynamic linking must be implemented separately for all supported ISAs
(x86, Arm, PowerPC, RISC-V, etc.), which makes the framework complicated and 
error-prone. For example, if the \ifunc{} library uses OpenMP to do
multi-threaded computation and is compiled by GCC, the compiler-generated
work-sharing functions need special treatment to be relocated correctly. Another
example is that older versions of GCC compiles C11 atomic operations provided by
\texttt{<stdatomic.h>} to special outlined functions. Bodies of these functions
are inserted by the linker, not visible in GCC's assembly output. Therefore, the
assembly patcher could not change how they are relocated and the \ifunc{}
crashes. The issues above originate from the fact that \cc{}~\cite{grodowitz2021cc}
is sending compiled machine code that is already linked. If we instead go to the
other end of the spectrum, send C source code from the source process, and
compile it on the target, it would be as portable as it gets, but the overhead
in terms of compilation time and size of the dependencies (headers, C compiler)
would be substantial.

\subsection{The New Approach - Bring the Bitcode}\label{sec:bringbitcode}
\begin{figure}[h]\includegraphics[width=0.9\columnwidth]{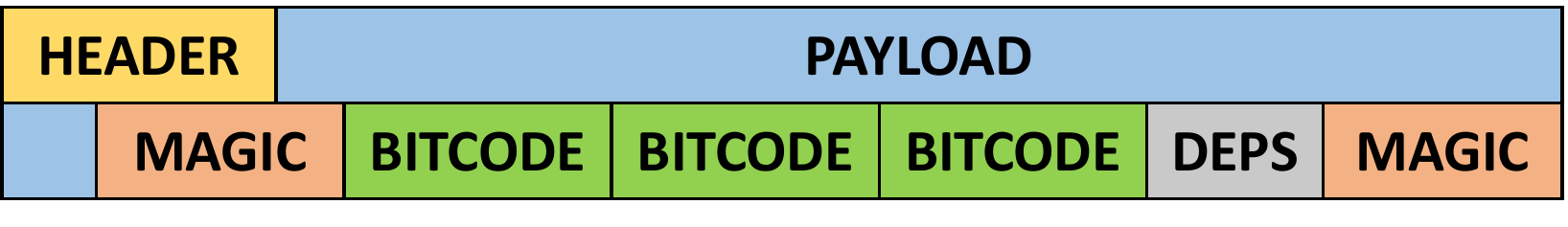}
 \centering
 \caption{Message frame of bitcode-based \ifuncs{}. \texttt{HEADER} carries 
 information that describes that type and format of the message, \texttt{MAGIC}
 field is used to discover and identify message delivery, \texttt{BITCODE}
 fields correspond to \fatir{}, and \texttt{DEPS} describe dependencies for the
 bitcode.}
 \label{fig:ifunc_msg_bc}
\end{figure}

LLVM is a popular framework for building compilers, it has been ported to many
different ISAs and is highly modular. LLVM provides a unified intermediate
representation (IR) that is capable to represent the semantics of high-level
programming languages, as well as a mature back-end compiler that transforms the
IR into efficient machine code. Today, many language implementations target the
LLVM IR so that they can focus on front-end optimizations while still getting
good portability across different ISAs. Clang, Julia, Rust, and Haskell are
among the most successful LLVM-based language implementations.

One key advantage of LLVM for this project is that it is natively a
cross-compiler, which means that generating IR/machine code for different ISAs
can be as simple as passing a single different flag to the same compiler program.
Since symbol resolution and linking happen downstream of LLVM IR generation,
compiling the source code to IR only needs the header files of the dependencies,
sans the special linkers and C runtime files required by GCC cross-compilation.
This feature of LLVM makes it very suitable for \ccc{}, which needs to run on
heterogeneous systems that have CPUs of different ISAs. Lastly, LLVM provides
JIT compilation for its IR through the ORC interface (ORC-JIT). ORC-JIT allows
the user to compile, link, and execute LLVM IR entirely in the memory, without
having to save the generated machine code to the disk and load it later. 

For these reasons, we have decided to implement a new \ifunc{} backend based on
LLVM. Instead of shipping binary code loaded from a dynamic library, we ship the
binary representation of the \ifunc{} library's LLVM IR, referred to as the
bitcode. When the target process receives a bitcode-based \ifunc{}, the \com{}
runtime will use LLVM ORC-JIT to compile the bitcode into machine code. This
compilation step does take some time, but the generated machine code is stored
in a LLVM internal buffer, which stays alive until the \ifunc{} is de-registered.
This means that subsequent \ifunc{} messages of the same type do not need to
repeat the compilation step and will be executed without delay.

Symbol resolution is also handled by LLVM in this new implementation. While the
LLVM IR does not contain information about shared library dependencies, ORC-JIT
does provide an interface to load shared libraries and perform run-time symbol
resolutions. In this new implementation, the user should provide a text file
\texttt{foo.deps} that lists all the dynamic library needed by \ifunc{} foo.
For example, OpenMP applications need \texttt{libomp.so} and OpenSSL
applications generally need \texttt{libcrypto.so}. This list of shared libraries
will also be shipped to the target process so that the \ccc{} runtime can load
them before invoking the main function of the \ifunc{}.

Since LLVM IR is ISA-dependent, an \ifunc{} message should contain bitcode for
all the ISAs it intends to run on. Therefore, the \com{} toolchain will generate
bitcode files for all the targets supported by the toolchain's Clang compiler.
These \texttt{.bc} files are available on the source process and are identified
by its target triple (e.g. \texttt{x86\_64-pc-linux-gnu}). When the \ifunc{} is
registered, all the bitcode files will be packed into a bitcode archive and for
the rest of the paper we reference it as a \fatir{}. The \fatir{} is shipped
with the payload and list of bitcode dependencies, as shown in Figure~\ref{fig:ifunc_msg_bc}.

The target process extracts the bitcode that matches its local target
architecture and compiles it with LLVM ORC-JIT. If we know the exact $\mu$Arch
of the target process's CPU and would like to optimize for it, we could provide
this information to Clang when we build the \ifunc{} bitcode. Then, on the target
process, ORC-JIT can instruct LLVM code-generation back-end to emit machine code
specialized for the $\mu$Arch of the CPU it is running on. In our experiments,
ORC-JIT running on Fujitsu A64FX was able to produce ARM LSE atomic instructions
and SVE vector instructions, using bitcode generated on an Intel Xeon CPU. Doing
the experiment the other way around, we have successfully generated x86 atomic
instructions and AVX2 vector instructions. This means that for performance
critical parts of the \ifunc{} that benefit from vectorization and/or atomics,
the bitcode implementation can reach the same level of performance as
binary-based \ifuncs{}.

One additional advantage of bringing in LLVM is that we can support \ifunc{}
libraries written in other high-level languages like Julia, as more and more
programming languages are using the LLVM compiler framework. We will discuss the
Julia integration in Section~\ref{subsec:julia}.

However, the LLVM bitcode based implementation does have a drawback: it pulls in
a portion of LLVM as a dependency. Although this portion is much smaller than a
full C compiler, it still is a 57 MiB file for the shared library version or
increases the size of the application binary by around 23 MiB if linked
statically. If the target system does not have a modern C++ compiler that could
build LLVM, or is extremely space-constrained, this could be an issue.

\subsection{Code Execution and Caching}\label{subsec:execandcache}

When the user creates an \ifunc{} message, the \ccc{} runtime packs the message
header, the payload, and the binary or bitcode into a single continuous block of
memory, called the message frame, as shown in Figures~\ref{fig:ifunc_msg_so}
and~\ref{fig:ifunc_msg_bc}. The receiving process of the \ifunc{} message first
registers the code section of the message to obtain an in-memory executable, and
then invokes the entry function with the payload and the target pointer.

While this process is relatively straightforward, the size of the binary or
bitcode shipped in an \ifunc{} message is a non-trivial source of communication
overhead. For example, even if the \ifunc{} is as simple as increasing a counter
on the target process, compiling the \ifunc{} library with high optimization
levels like \texttt{-O3} can increase the size of the shipped binary code from
65 bytes to 90 bytes, and the increase is even more significant for more complex
libraries. The overhead is particularly obvious for the new bitcode
implementation, as it must ship a \fatir{} that supports multiple ISAs of the
participating processes. For the previously mentioned counter-increasing \ifunc{},
its bitcode archive that supports both x86\_64 and AArch64 processors is around
5 KiB. Shipping such a large amount of extra data could have a significant
negative impact on communication latency and message rate for small messages,
even on high-performance \acrshort{rdma} interconnects.

\begin{figure}[h]
 \centering
 \includegraphics[width=0.75\columnwidth]{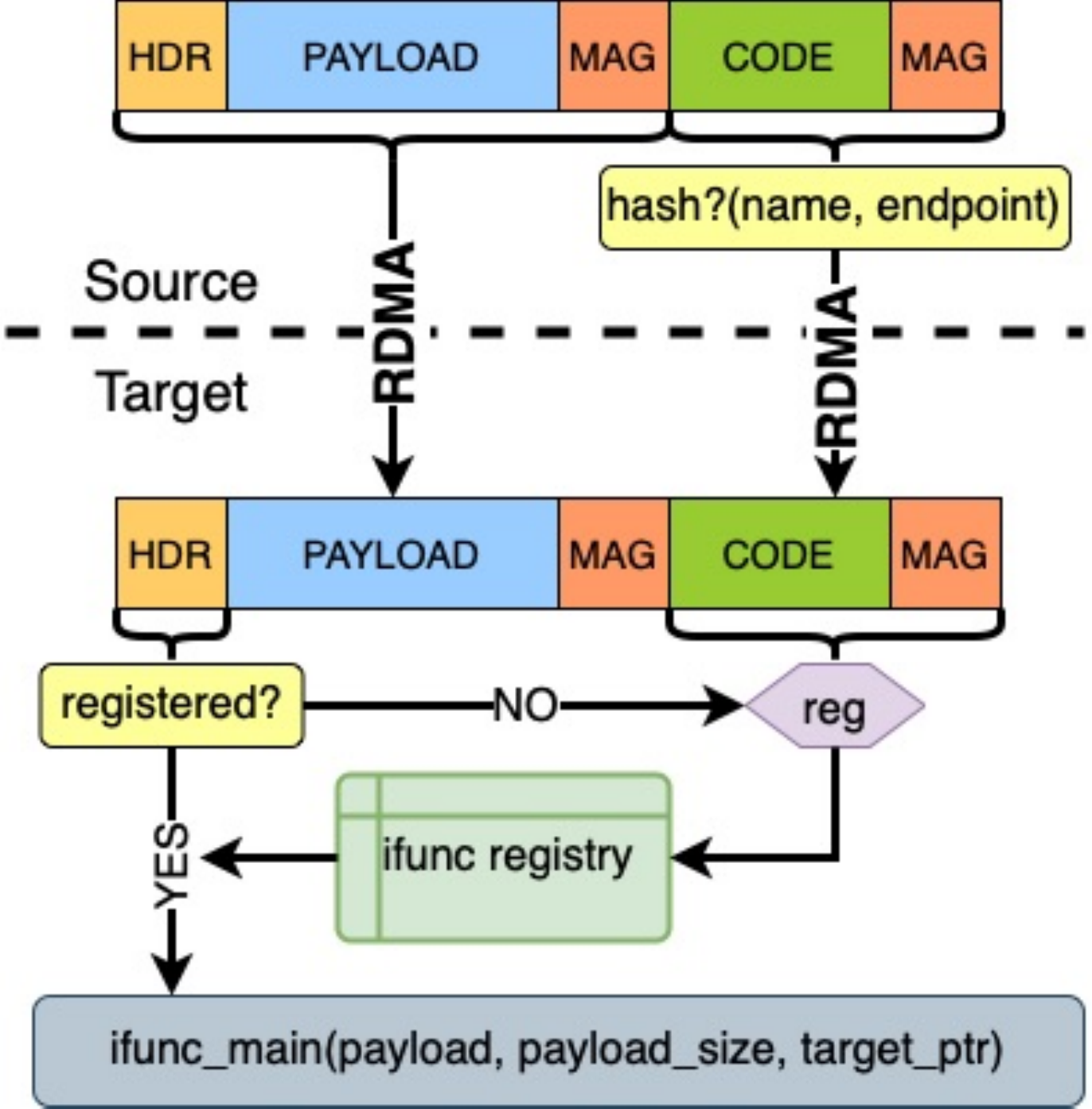}
 \caption{\ifunc{} caching mechanism}
 \label{fig:ifunc_caching}
\end{figure}

To address this issue, we have added code caching mechanism for both \ifunc{}
implementations. To keep the simplicity of the API, all user-created \ifunc{}
messages still contain both code and data, and the UCX runtime will perform
caching fully transparent to the user. The process is demonstrated in
Figure~\ref{fig:ifunc_caching}.

For both the binary and the bitcode implementation, the code is placed at the
last section of the \ifunc{} message, in between two signals (\texttt{MAGIC})
bytes. The idea here is that we always construct a full message, but skip
sending the code section if we know the target process has already seen this
type of \ifunc{} at least once. Note that we do not need two separate calls to
UCX communication routines to send the entire message, the message frame is
contiguous in memory, and we control what to send by simply passing different
message size arguments to the UCP PUT interface.

When the target process receives an \ifunc{} message, the \ccc{} runtime first
looks at the message header and see if this \ifunc{} has already been registered
on the local process. If this is the first time this process receives an \ifunc{}
of this type, then we assume we have received a full message that contains the
code section. The runtime will then automatically register this \ifunc{} and
copy the code section to a side buffer for execution. For binary-based \ifuncs{},
the side buffer will have proper alignment and execution privilege, and the
runtime will patch its GOT before starting the execution. For bitcode-based
\ifunc{}s, the runtime will save the bitcode archive to the side buffer, create
a LLVM ORC-JIT instance with the bitcode that matches the local process's target
architecture, and start execution.

If the target process has already registered the \ifunc{} before, then it
assumes the message was truncated and so only waits for the delivery of the
payload section. For both binary and bitcode \ifunc{}s, we have stored the
pointer to their \texttt{main} functions during registration, so the runtime
invokes the \ifunc{}'s \texttt{main} function immediately once the payload is
fully delivered.

When the source process sends an \ifunc{} message, the \ccc{} runtime first
checks a hash table to see if it has sent an \ifunc{} message of this particular
type to the specified UCP endpoint before. If not, then the endpoint is added to
the hash table and the entire message is sent. If the UCP endpoint is already in
the hash table, we know the target has already cached the code for this type of
\ifunc{}, so the \ccc{} runtime will only send the message up to the second last
signal byte, skipping the code section and the trailer signal byte. This hash
table lookup adds a bit of overhead, but it's very effective in reducing the
message size and improving performance. Note that the \ifunc{} message is never
modified in this process, as the user might want to send it to another process
later.

\subsection{Integration with Julia Language}\label{subsec:julia}
The Julia programming language uses LLVM for its JIT ahead-of-time compilation
approach. It currently has only limited tooling for static compilation and very
limited support for cross-compilation.

Since Julia is a dynamic programming language there exists uncertainty about the
function call-graph of a program. Julia uses an abstract-interpretation based
approach where on the first call of a method with a given type-signature, the
types of variables are inferred, and the call-graph is (if possible) resolved
ahead of time. If a user method contains uncertainty -- colloquially referred to
as type-instability -- a dynamic call is inserted. Dynamic calls will be resolved
at runtime through a dynamic dispatch, where the runtime variables are examined,
and the correct method is selected.

We take advantage of the GPU compiler infrastructure (\texttt{GPUCompiler.jl}~\cite{Besard2017-it, Besard2019-zu})
that was developed to allow the user to offload Julia functions to GPGPU
accelerators. \texttt{GPUCompiler.jl} collects the statically reachable
call-graph of a Julia function into one LLVM IR module. It disallows dynamic
dispatch, as well as accessing of runtime managed data such as global variables
or Julia runtime functions. This provides us with a subset of the Julia language
that is highly performant and can be optimized well with LLVM. 

We then pass LLVM IR module generated by \texttt{GPUCompiler.jl} to \ccc{} and
use \ifunc{} polling function to invoke them on the target processing unit. One problem
that arose was that Julia does not currently support cross-compilation and for
\ccc{} we compile the \ifunc{} on the target systems and cache the resulting
bitcode as a file.

In future work we could relax the limitation on runtime access and allow
full-fledged support for global variables, memory allocation (GC) and the task
runtime, but this and improved cross-compilation both necessitate improvements
to the Julia compiler.

The Julia implementation builds on top of \texttt{UCX.jl}, a set of Julia
bindings for UCX, and we developed new bindings for the API introduced by our
framework. Besides the use of \texttt{GPUCompiler.jl} to exfiltrate LLVM IR to
be used on the target machines, the workflow is identical to the code flow
described in Sections~\ref{sec:bringbitcode} and~\ref{subsec:execandcache}. This
allows our framework to consume both \ifuncs{} written in Julia or C, by either
a Julia or C based applications.

%% file: content/experiments.tex
\section{Experimental Design}\label{sec:exp-design}
\subsection{Benchmark Design}
To further characterize our \ccc{} framework, we created two benchmarks that use
the \ifunc{} API. One of these is the \acrfull{tsi} \ifunc{}, and the other one
is the \acrfull{dapc} miniapp. Each one of the benchmarks implements three
different modes of code execution: \am{}, \ifunc{} with binary code
representation, and \ifunc{} with bitcode code representation. The \am{} mode
assumes that the binary code is already compiled and present on the target
machine. The \am{} request only transfers payload data and an index pointing to
the function in a pointer table. The \am{} mode is only used as an evaluation
baseline to calculate the overheads of transmitting and executing \ifuncs{}.
The \ifunc{} binary and bitcode representations are implemented as described in
Sections \ref{sec:ifunc-binary} and \ref{sec:bringbitcode}.

\subsection {\acrfull{tsi}}\label{sec:tsi}
To measure the overheads of \ifunc{} transmission, \jitting{}, and 
execution to a target system, we implemented a simple benchmark that measures 
message rate and latency for an \ifunc{} that just increases a counter on a
target machine. The increment function is injected to a remote machine with a
1-byte payload and 5159 bytes of bitcode. Using this benchmark, we measured the
baseline overheads for moving compute and data to a remote system.

\subsection{\acrfull{dapc}}\label{sec:dapc}
To provide a more realistic evaluation of the \ifunc{}, we developed
code that represents a new class of \acrfull{xrdma} operations. The basic idea
behind \acrshort{xrdma} is an implementation of a complex \acrshort{rdma}
operation where a user injects user-defined code and data into a remote
processing element on a target DPU or server. The injection operation can modify
remote memory and issue new remote memory operations using \ifunc{}. To
demonstrate these new capabilities, we implemented the \acrfull{dapc} miniapp. 
\acrshort{dapc} implements a workload that resembles pointer chasing across
multiple distributed machines. In our implementation, the pointer table is
evenly distributed across the servers and the pointer chasing part of the
algorithm is implemented using the \ccc{} framework. Each step of the pointer 
chase the algorithm has the flexibility of calling itself recursively, sending 
itself as an \ifunc{} or creating another \ifunc{} with new logic. As a result,
we call this algorithm adaptive as it changes what it does as it proceeds based
on context and input parameters. The data (or pointer tables) are evenly spread
among the server machines into shards of the same size and the entries are
indexed using the server number first. These machines are waiting for incoming
\ifuncs{}. When the client machine is ready to start a pointer chase operation,
it creates a \acrshort{xrdma} \textit{Chaser} operation using an \ifunc{} with
the following fields:
\begin{itemize}
 \item \textit{Address}: the first element to be accessed
 \item \textit{Depth}: the depth of the pointer chase
 \item \textit{Destination}: the node ID of the client/requester
\end{itemize}
When the \ifunc{} message is ready to be sent, the client sends it to
\textit{any} of the servers. When a server receives a \textit{Chaser} \ifunc{},
it calls it, passing local information like the address of the pointer table.
The \ifunc{} determines if it is running in the server where the entry is
located. If it is running in the wrong server, it forwards an \acrshort{xrdma}
\textit{Chaser} \ifunc{} to the correct server. Once running in the right place,
the \ifunc{} loads the entry. If this entry is the final one in the pointer
chase, the \ifunc{} sends the result back to the client. If this entry is the
\textit{Address} of the next entry, then the \ifunc{} determines if the entry is
local or remote. If the entry is local, the \ifunc{} calls itself recursively.
Otherwise, the \ifunc{} is forwarded to the correct server.

To return the result to the requester, we created the simple \acrshort{xrdma}
\textit{ReturnResult} operation. All it does is return the result to the
requester. In the case of an \acrshort{xrdma} deployment on a DPU, all the
\ifunc{} waiting and recursive \ifunc{} invocations are completely offloaded to 
the Arm cores on the DPU. 

\subsection{\acrfull{gbpc}}\label{sec:gbpc}
The \acrshort{dapc} operation has a \textit{baseline} mode, the \acrfull{gbpc}.
This mode performs the pointer chase using UCX GET semantics to remotely load
the entries from the servers. It uses an iterative approach to issue remote UCP
GETs until the final entry of the pointer chase is found. The code for this
approach is simpler than the \ifunc{} code, but the client must do all the
work. \acrshort{gbpc} is implemented as a special case on the \acrshort{dapc}
code. 

\subsection{Test Configurations}\label{sec:configs}
We evaluate \acrshort{dapc} and \acrshort{gbpc} by performing a parameter sweep
across \emph{depth} and \emph{number of servers}. This allows us to study the
performance characteristics of \ifunc{} on several regimes. For \acrshort{dapc},
the partitioning of the data is refined as the number of servers increases, and,
thus, the fraction of cross-server communication for \acrshort{dapc} rises. We
implemented the setup both in C and Julia. The Julia implementation was kept as
close as possible to the original C implementation.

\subsection{Testbed Platforms} \label{sec:testbeds}
For our evaluation we used two different clusters. The first one is the \ookami{}
HPE Cluster at Stony Brook University. The \ookami{} cluster is based on the
Apollo 80 HPE platform, and the system has 174 Fujitsu A64FX FX700 compute nodes
based on the ARMv8 CPU architecture. The A64FX SoC has a total of 48 cores
clocked to 1.8/2.0 GHz with 32GB HBM memory. The cluster is connected with
InfiniBand network using Mellanox ConnectX-6 100 Gb/sec dual-port host channel
adapters. The system is running CentOS Linux 8.4.2105 with Linux Kernel
4.18.0-305.25.1.el8\_4.aarch64, GCC 11.3.0, and MOFED 5.4-3.1.0.0 The second
system is the \thor{} cluster at the HPC Advisory Council. \thor{} is based on
Dell PowerEdge R730 platform with Dual Socket Intel Xeon 16-core CPUs E5-2697A
clocked to 2.60 GHz. The system has 36 nodes each equipped an with Arm-based
(Cortex-A72) NVIDIA \acrfull{bf2} 100Gb/s InfiniBand DPU adapter. The system is
running Red Hat Enterprise Linux 8.5 with kernel 4.18.0-348.12.2.el8\_5.x86\_64,
GCC 8.5.0, and MOFED 5.5-1.0.3.2. On all systems we used \ccc{} implementation
based on LLVM 13.0.1, Julia 1.8.0-beta3, and OpenUCX master
(\texttt{508e766d}). It is important to point out that we got \ccc{} framework
running on multiple other systems, including computational storage devices
connected over Ethernet network. For this paper we selected what we believe are
the most interesting systems.

%% file: content/eval.tex
\section{Experimental Evaluation and Analysis}\label{sec:eval}

\subsection{Bitcode \ifunc{} Overhead Breakdown}\label{sec:breakdown}
One of the first questions we wanted to answer was what the overheads were of
using our \ccc{} framework to send and run \ifuncs{}. To measure these
overheads, we used the kernel described in Sec.~\ref{sec:tsi} as the code we
wanted to execute on the target system. We wrote a benchmark that measures the 
latencies of running the \acrshort{tsi} kernel in \am{}, cached \ifunc{} and 
uncached \ifunc{} modes between a pair of systems. In addition to measuring the
latency of running the \acrshort{tsi} code, we measured the latency of sending 
cached \ifuncs{} (26 bytes), \ams{} (33 bytes), and uncached \ifuncs{} (5185 bytes).
With all these measurements, we estimate what the overheads of each of
the steps of sending and running an \ifunc{}. The four main steps of issuing
an \ifunc{} to a target system are:
\begin{itemize}
 \item Transmission. The source sends the bitcode and data using a \texttt{Put} operation.
 \item Lookup. The target checks if the bitcode has already been \jitted{}
   by LLVM and cached by \ccc{}. If not cached by \ccc{}, the target checks if 
   the bitcode has already been \jitted{} by LLVM.
 \item \jitting{}. If the bitcode is not already \jitted{} and cached, the target's
   LLVM JITs the bitcode and caches the binary generated. This step performs
   the dynamic linking of dependencies.
 \item Execution. The target runs the \jitted{} binary.
\end{itemize}

We used the following equations to generate the overhead tables below. In these
equations, we abbreviated \am{} to \textit(AM), cached \ifunc{} to \textit{C},
and uncached \ifunc{} to \textit{U}. The \textit{total}, \textit{trans},
\textit{JIT}, and \textit{L+E} abbreviations correspond to total, transmission,
JIT, and lookup-and-execution overheads, respectively.

\begin{align}
    U_{total} &= U_{trans} + U_{JIT} + U_{L+E} \label{eq:nc-tot} \\
    C_{total} &= C_{trans} + C_{L+E} \label{eq:c-tot} \\
    AM_{total} &= AM_{trans} + AM_{L+E} \label{eq:am-tot}
\end{align}

It is important to point out that 
LLVM's ORC-JIT caches observed code symbols. Because of this internal caching,
we observed no overhead for \jitting{} in the benchmarks that re-send the same \ifunc{}. When \ccc{} invokes JIT operation for the \ifunc{}, LLVM has to do minimal work since it looks up the \ifunc{} from previous JIT invocations.
To get an accurate estimate of JIT overheads without LLMV ORC caching we created 
a separate benchmark to measure the \jitting{} duration without the caching and listed this time in the table (JIT), but did not add it to the total time.

On~\autoref{table:fujitsu-breakdown}, we list how long each step takes for all
three modes when tested on \ookami{} machines. The \am{} and cached bitcode
columns do not include a value for \jitting{} because these operations do not need
to perform \jitting{}. We estimated that \jitting{} for the uncached bitcode took
approximately \SI{6.59}{\milli\second}, completely dominating the transmission and
execution times. We can also observe the toll sending bitcode takes on transmission
time. The cached \ifunc{} message is just 26B, while the bitcode sent with the
uncached version adds 5,159 bytes, making the message 5,185 bytes long. Sending the
uncached bitcode takes almost double the time it takes to send the cached \ifunc{},
taking \SI{2.67}{\micro\second} and \SI{5.12}{\micro\second}, respectively. Overall,
the uncached bitcode takes 91\% longer to get sent and executed.

On the \ookami{} system, the cached \ifuncs{} perform pretty well, with message
transmission and execution taking about 3\% longer than \ams{}.

\autoref{table:bf2-breakdown} shows the time breakdown for each mode when tested
between two \acrshort{bf2} systems from the \thor{} cluster. In this system,
\jitting{} the \acrshort{tsi} code takes \SI{4.50}{\milli\second}. Sending the
larger uncached \ifunc{} message takes 86\% longer than sending the cached
version, and the uncached version takes 88\% longer to finish end-to-end.

\begin{table}[t]
\begin{center}
\setlength\tabcolsep{4pt}
\begin{tabular}{|l|r|r|r|}
 \hline
 \rowcolor{lightgray} Stage & Active Message           & Uncached Bitcode           & Cached Bitcode           \\
 \hline{}
 Lookup+Exec                  & \SI{0.08}{\micro\second} & \SI{0.10}{\micro\second}   & \SI{0.05}{\micro\second} \\
 \ JIT                        & N/A                      & (\SI{6.59}{\milli\second}) & N/A                      \\
 \ Transmission               & \SI{2.50}{\micro\second} & \SI{5.02}{\micro\second}   & \SI{2.62}{\micro\second} \\
 \hline{}
 Total                      & \SI{2.58}{\micro\second} & \SI{5.12}{\micro\second}   & \SI{2.67}{\micro\second} \\
 \hline
\end{tabular}
\end{center}    
\caption{\ookami{} \acrshort{tsi} overhead breakdown}\label{table:fujitsu-breakdown}
\end{table}

\begin{table}[t]
\begin{center}
\setlength\tabcolsep{4pt}
\begin{tabular}{|l|r|r|r|}
 \hline
 \rowcolor{lightgray} Stage & Active Message           & \jitted{} Bitcode          & Cached Bitcode           \\
 \hline{}
 Lookup+Exec                & \SI{0.01}{\micro\second} & \SI{0.04}{\micro\second}   & \SI{0.01}{\micro\second} \\
 \ JIT                        & N/A                      & (\SI{4.50}{\milli\second}) & N/A                      \\
 \ Transmission               & \SI{1.87}{\micro\second} & \SI{3.45}{\micro\second}   & \SI{1.85}{\micro\second} \\
 \hline{}
 Total                      & \SI{1.88}{\micro\second} & \SI{3.49}{\micro\second}   & \SI{1.86}{\micro\second} \\
 \hline
\end{tabular}
\end{center}    
\caption{\thor{} \acrshort{bf2} \acrshort{tsi} overhead breakdown}\label{table:bf2-breakdown}
\end{table}

\begin{table}[b!]
\begin{center}
\setlength\tabcolsep{4pt}
\begin{tabular}{|l|r|r|r|}
 \hline
 \rowcolor{lightgray} Stage & Active Message           & \jitted{} Bitcode          & Cached Bitcode           \\
 \hline{}
 Lookup+Exec                & \SI{0.01}{\micro\second} & \SI{0.01}{\micro\second}   & \SI{0.02}{\micro\second} \\
 \ JIT                        & N/A                      & (\SI{0.83}{\milli\second}) & N/A                      \\
 \ Transmission               & \SI{1.55}{\micro\second} & \SI{3.58}{\micro\second}   & \SI{1.51}{\micro\second} \\
 \hline{}
 Total                      & \SI{1.56}{\micro\second} & \SI{3.59}{\micro\second}   & \SI{1.53}{\micro\second} \\
 \hline
\end{tabular}
\end{center}    
\caption{\thor{} Xeon \acrshort{tsi} overhead breakdown}\label{table:xeon-breakdown}
\end{table}

\begin{table}[t]
\begin{center}
\setlength\tabcolsep{4pt}
\begin{tabular}{|l|r|r|r|r|}
 \hline
 \rowcolor{lightgray} Method & Latency                  & Speedup                  & Message Rate      & Speedup                   \\
 \hline
 Active Message              & \SI{2.58}{\micro\second} & \multirow{2}{*}{-3.29\%} & 1,320,000 msg/sec & \multirow{2}{*}{26.44\%}  \\
 Cached Bitcode              & \SI{2.67}{\micro\second} &                          & 1,669,000 msg/sec &                           \\
 \hline
 Uncached Bitcode            & \SI{5.12}{\micro\second} & \multirow{2}{*}{91.39\%} &   405,300 msg/sec & \multirow{2}{*}{311.79\%} \\
 Cached Bitcode              & \SI{2.67}{\micro\second} &                          & 1,669,000 msg/sec &                           \\
 \hline
\end{tabular}
\end{center}
\caption{\ookami{} \acrshort{tsi} latencies and message rates}\label{table:fujitsu-values}
\end{table}

\begin{table}[t!]
\begin{center}
\setlength\tabcolsep{4pt}
\begin{tabular}{|l|r|r|r|r|}
 \hline
 \rowcolor{lightgray} Method & Latency                  & Speedup                  & Message Rate      & Speedup                   \\
 \hline
 Active Message              & \SI{1.88}{\micro\second} & \multirow{2}{*}{0.86\%}  &   974,000 msg/sec & \multirow{2}{*}{34.60\%}  \\
 Cached Bitcode              & \SI{1.87}{\micro\second} &                          & 1,311,000 msg/sec &                           \\
 \hline
 Uncached Bitcode            & \SI{3.49}{\micro\second} & \multirow{2}{*}{87.73\%} &   417,300 msg/sec & \multirow{2}{*}{214.16\%} \\
 Cached Bitcode              & \SI{1.87}{\micro\second} &                          & 1,311,000 msg/sec &                           \\
 \hline
\end{tabular}
\end{center}
\caption{\thor{} \acrshort{bf2} \acrshort{tsi} latencies and message rates}\label{table:bf2-values}
\end{table}

\begin{table}[b!]
\begin{center}
\setlength\tabcolsep{4pt}
\begin{tabular}{|l|r|r|r|r|}
 \hline
 \rowcolor{lightgray} Method & Latency                  & Speedup                   & Message Rate      & Speedup                   \\
 \hline
 Active Message              & \SI{1.56}{\micro\second} & \multirow{2}{*}{2.30\%}   & 6,754,000 msg/sec & \multirow{2}{*}{8.11\%}   \\
 Cached Bitcode              & \SI{1.53}{\micro\second} &                           & 7,302,000 msg/sec &                           \\
 \hline
 Uncached Bitcode            & \SI{3.59}{\micro\second} & \multirow{2}{*}{135.54\%} & 2,037,000 msg/sec & \multirow{2}{*}{258.47\%} \\
 Cached Bitcode              & \SI{1.53}{\micro\second} &                           & 7,302,000 msg/sec &                           \\
 \hline
\end{tabular}
\end{center}
\caption{\thor{} Xeon \acrshort{tsi} latencies and message rates}\label{table:xeon-values}
\end{table}

In this configuration, the \am{} and cached \ifunc{} take essentially the same
time to execute once received. Since the cached \ifunc{} message is slightly
smaller than the \am{} (26B versus 33B), the transmission time favors the cached
\ifunc{} (about 1\% faster).

\autoref{table:xeon-breakdown} shows the data collected and calculated from the 
\thor{} Xeon systems. \acrshort{tsi} \jitting{} takes \SI{0.83}{\milli\second}.
In this configuration, not sending the \ifunc{} code provides an overall speedup of
2.3$\times{}$. These systems launch the cached \jitted{} code 2\% faster
than they launch the \am{} code.

Our main take away from these experiments is that \jitting{} incurs an expensive
one-time cost. Since our \acrshort{tsi} kernel only increments a memory value, it
executes so quickly, that its execution time is negligible. The dominating cost
of running this simple \ifunc{} is sending it to the target.

In this section, we do not include binary \ifunc{} numbers because the effect of
caching on execution is negligible since the code arrives ready to be executed.
The only benefit of caching for binary \ifuncs{} is on the transmission side,
where the uncached version of \acrshort{tsi} is 75 bytes compared to the 26-byte
cached version.

\subsection{Bitcode \ifunc{} Latency and Message Rate Overheads}

As part of investigating the overheads of our bitcode \ifuncs{}, we also tested
the message rate achievable when sending and executing as many \acrshort{tsi}
kernels as possible from one machine to another. We tested these using \am{},
cached \ifunc{} and uncached \ifunc{}. We listed these results in the following
tables, alongside the latency results from~\autoref{sec:breakdown}.

\autoref{table:fujitsu-values},~\autoref{table:bf2-values},
and~\autoref{table:xeon-values} show the results of the latency and message rate
of executing the \acrshort{tsi} kernel using \ams{}, uncached \ifunc{} and
cached \ifunc{}. Caching the \jitted{} code drastically improves latency and
message rate for all systems tested. Depending on the configuration, we observed
latency reductions of 84\%-144\% and message rate improvements of 214\%-311\%.

When comparing cached \ifuncs{} with \ams{}, \ifuncs{} consistently exhibit a
higher message rate (8\%-34\%). In the \thor{} configurations
(\autoref{table:bf2-values} and \autoref{table:xeon-values}), cached \ifunc{}
latency is better than that of \am{}. Only in the \ookami{} configuration
(\autoref{table:fujitsu-values}) we observe a worse latency for \ifuncs{}.

Overall, cached \ifuncs{} exhibit between slightly worse (up to 3\% higher
latency) to significantly better performance (up to 34\% higher message rate)
than \ams{}. \ifuncs{} provide the dynamic programmability that \ams{} cannot.

\subsection{Pointer Chase Performance Depth Sweep}\label{sec:eval-depth}
To better understand the performance of \ifuncs{} in a more realistic scenario,
we tested the \acrfull{dapc} \ifunc{} implementation (described
in Sec.~\ref{sec:dapc}) and compared it to the \am{} implementation and to
\acrfull{gbpc}. We ran a range of pointer chasing depths, from 1, scaling in
powers of 2, all the way to 4096. For these sweep tests, we ran on the \ookami{}
and \thor{} clusters using a variety of configurations. For the sake of space,
we only present a very small subset of the results obtained, but these are
representative of the trends we observed across all configurations tested.

\begin{figure}[t!]\includegraphics[width=0.90\linewidth]{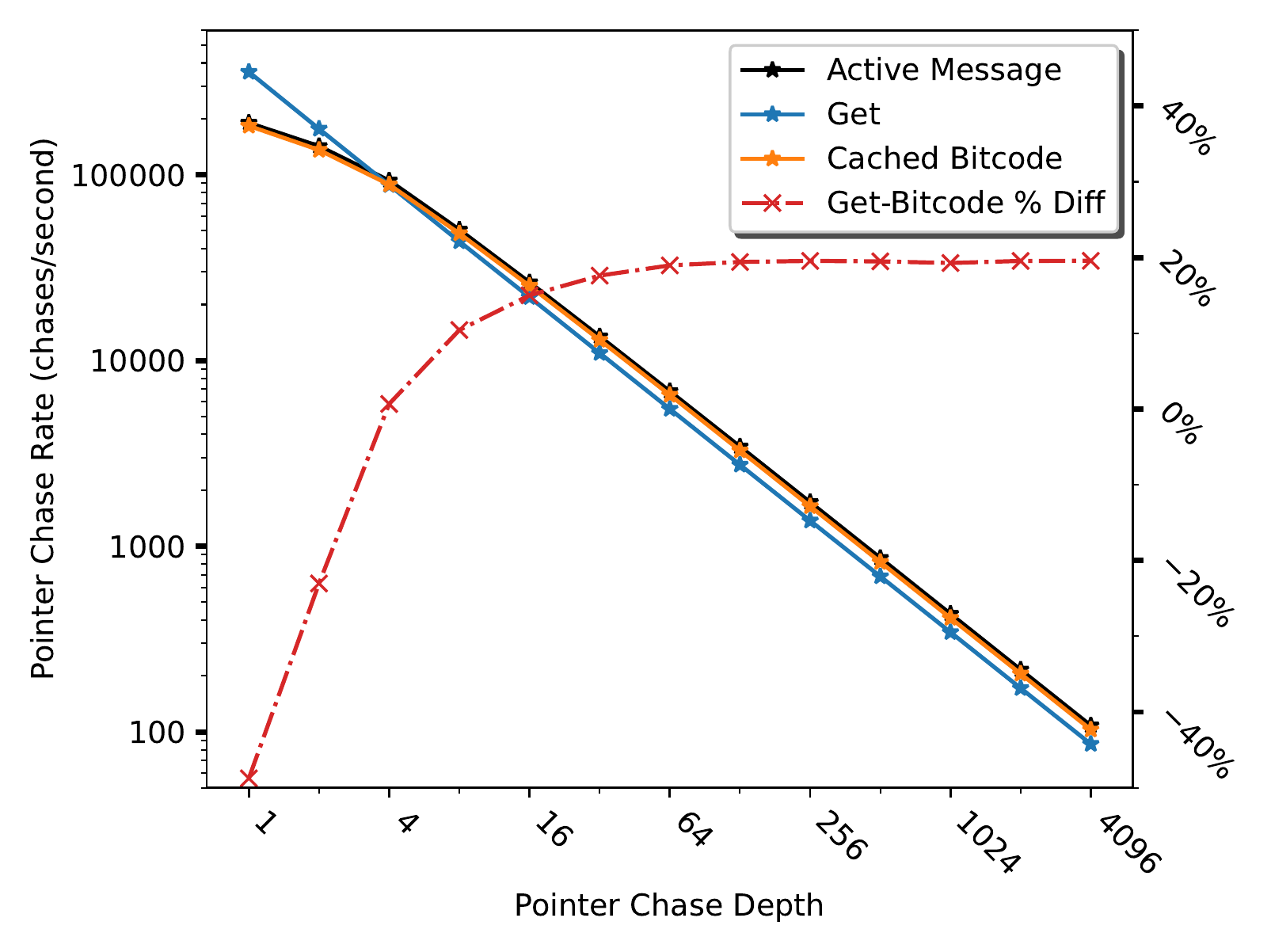}
 \centering
 \caption{\thor{} 32-Server; \textbf{C/C++} (Xeon Client and \acrshort{bf2} Servers): \acrfull{dapc}}
 \label{fig:xeon-bf-depth}
\end{figure}

\begin{figure}[t!]\includegraphics[width=0.90\linewidth]{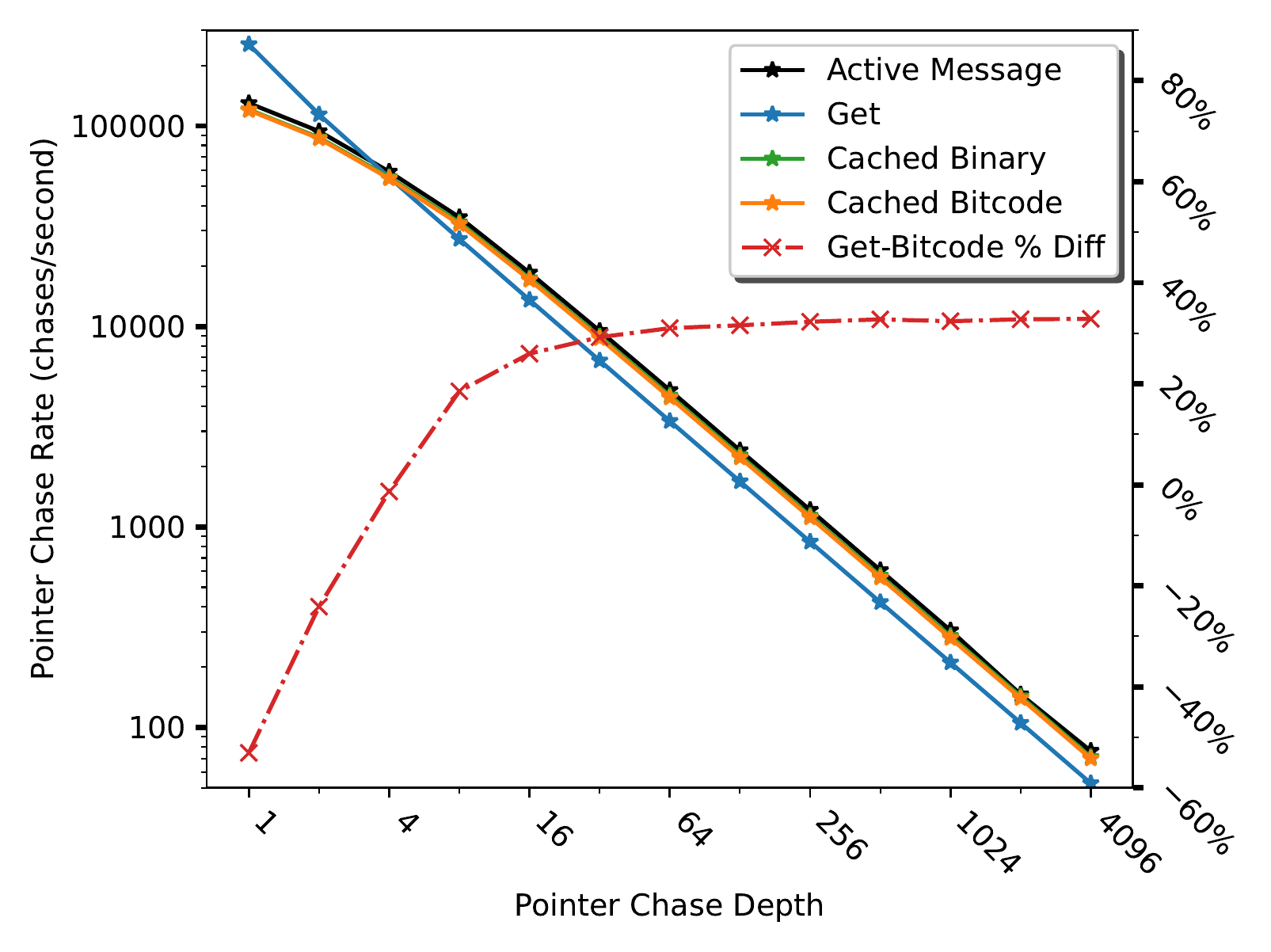}
 \centering
  \caption{\ookami{} 64-Server; \textbf{C/C++}: \acrfull{dapc}}
 \label{fig:fujitsu-depth}
\end{figure}

\begin{figure}[t!]\includegraphics[width=0.90\linewidth]{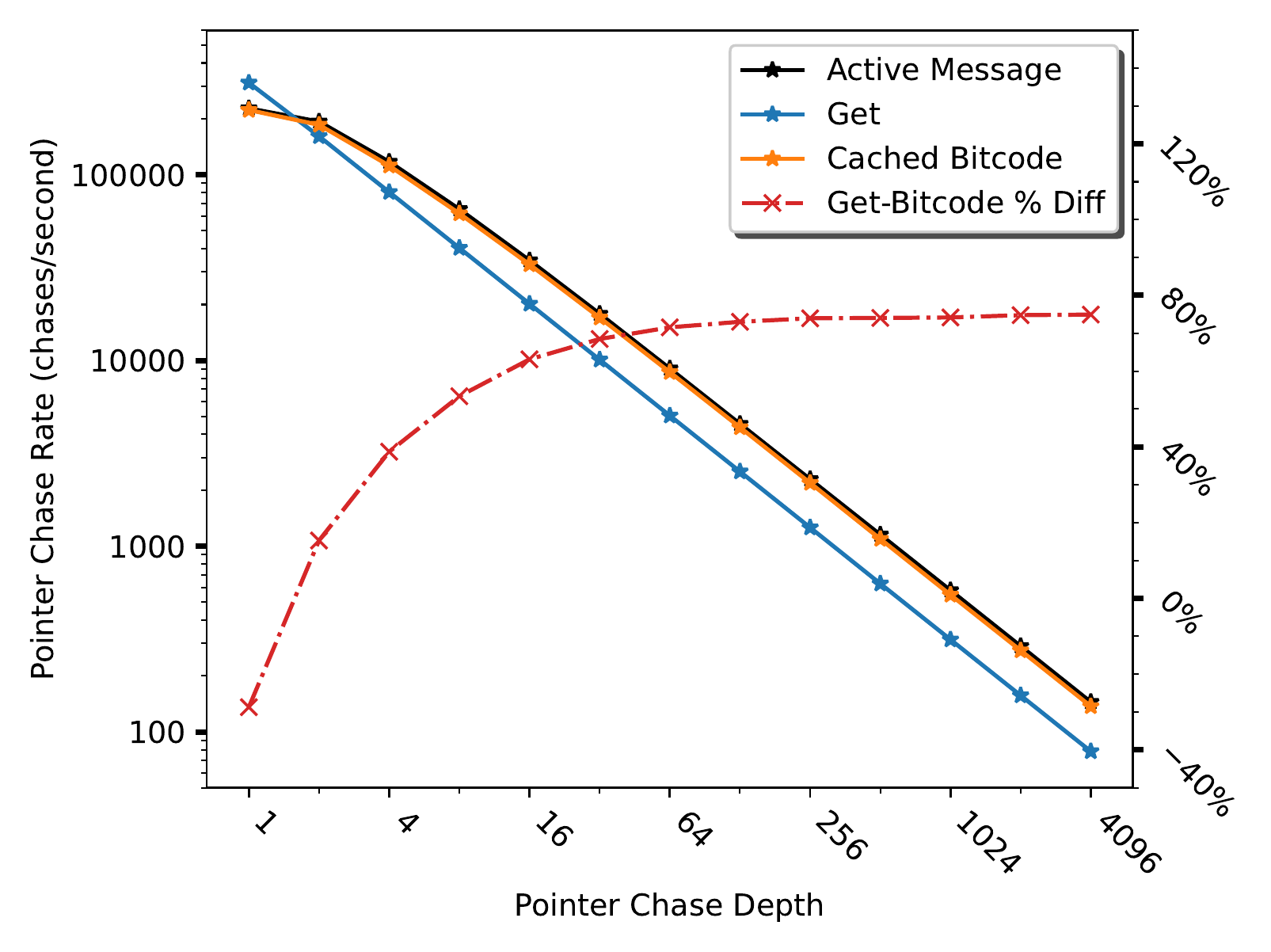}
 \centering
 \caption{\thor{} 16-Server; \textbf{C/C++} (Xeon Client and Servers): \acrfull{dapc}}
 \label{fig:xeon-xeon-depth}
\end{figure}

\begin{figure}[t!]\includegraphics[width=0.90\linewidth]{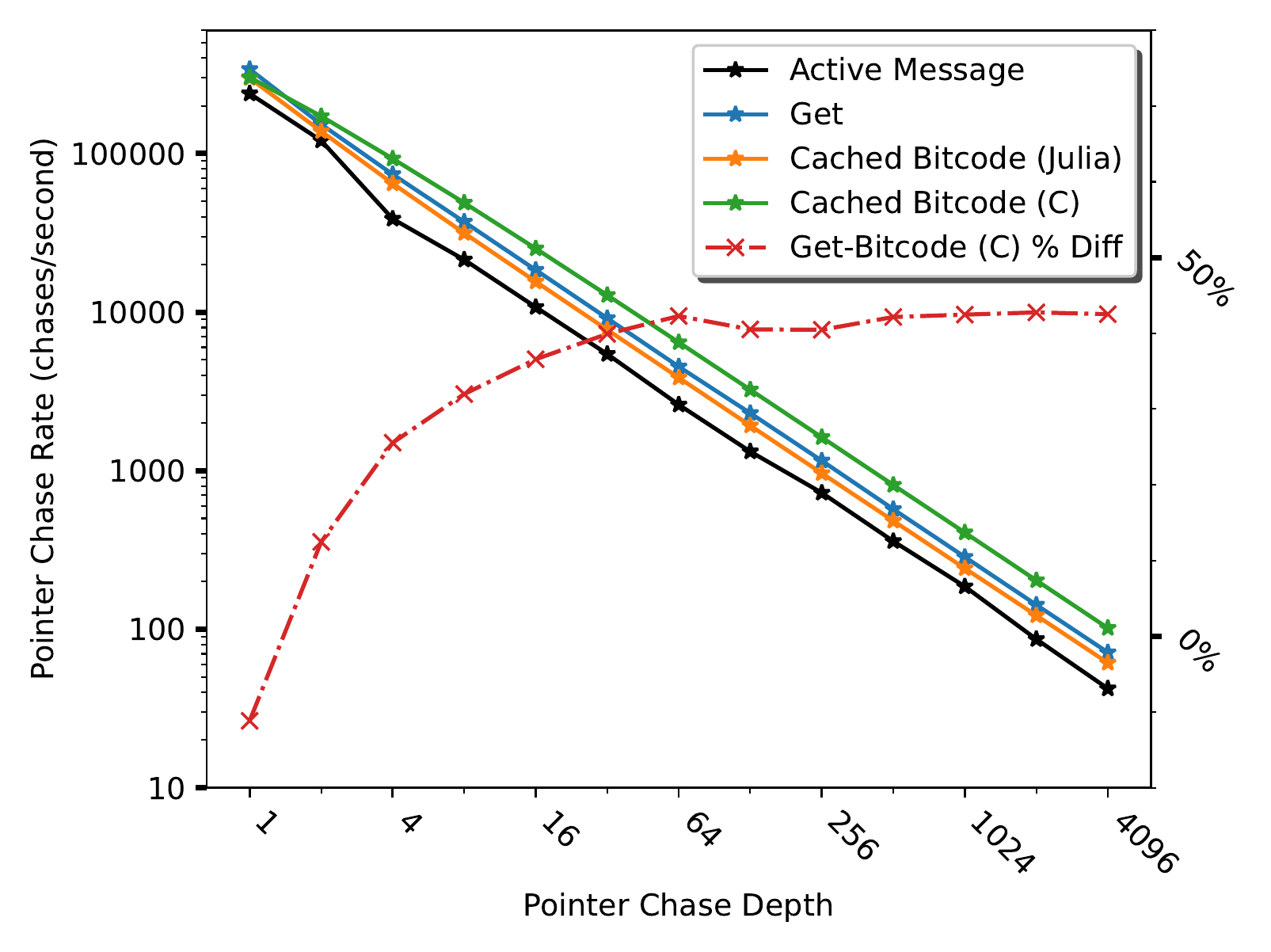}
 \centering
 \caption{\thor{} 32-Server; \textbf{Julia} (Xeon Client and \acrshort{bf2} Servers): \acrfull{dapc}}
 \label{fig:xeon-bf-depth-julia}
\end{figure}

Figures \ref{fig:xeon-bf-depth}, \ref{fig:fujitsu-depth},
\ref{fig:xeon-xeon-depth}, and \ref{fig:xeon-bf-depth-julia}
show how many pointer chases can complete per
second for all our system and combination of C/C++ and Julia implementations.
For the \acrshort{dapc} C-based implementation we observe nearly identical
behavior. The cached bitcode \ifunc{} implementation performs on pair with
\ams{} and cached binary implementation on Arm-based platforms. The bitcode
implementation also outperforms \acrshort{gbpc} (Get) implementation by up to 20\% on \thor{} with \acrshort{bf2}
Servers (\autoref{fig:xeon-bf-depth}), by up to 30\% on the \ookami{} system
(\autoref{fig:fujitsu-depth}), and by up to 75\% on the \thor{} with Xeon
servers (\autoref{fig:xeon-xeon-depth}).

As depth grows, the gap between each of the lines stabilizes to a constant value.
The gap between each line pair represents the constant speed difference between
pointer lookup and traversing using each one of the methods.

\subsection{Pointer Chase Performance Scalability}\label{sec:eval-scale}

We used the data collected to understand what happens to pointer chase rate when
scaling out to more servers. We took the 4096-depth datapoints for all modes
tested and plotted them against number of servers.

\begin{figure}[h]\includegraphics[width=0.90\linewidth]{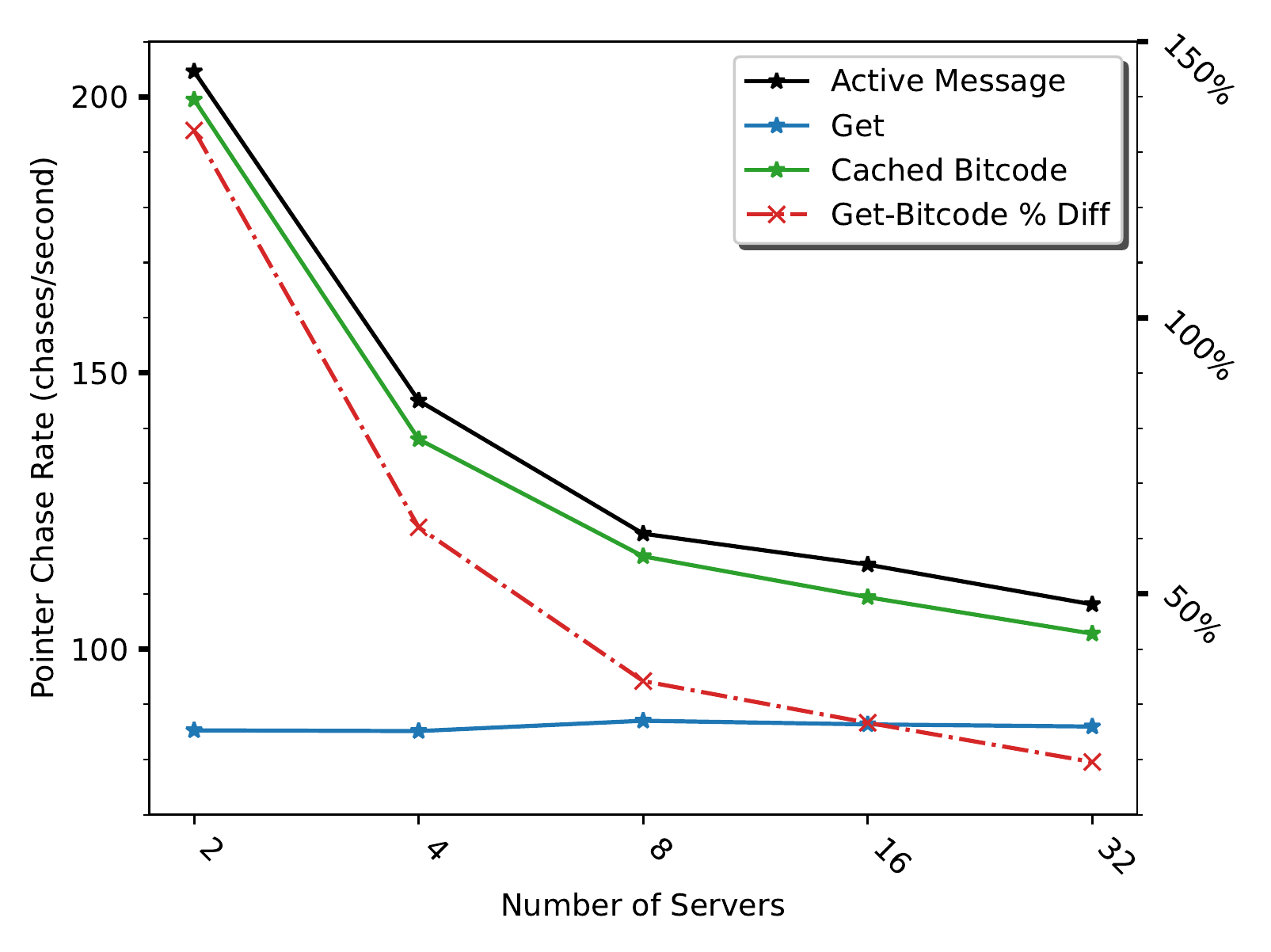}
 \centering
  \caption{\thor{} 4096-Chase-Depth; \textbf{C/C++} (Xeon Client and \acrshort{bf2} Servers): \acrfull{dapc}}
 \label{fig:xeon-bf-scale}
\end{figure}

\begin{figure}[h]\includegraphics[width=0.90\linewidth]{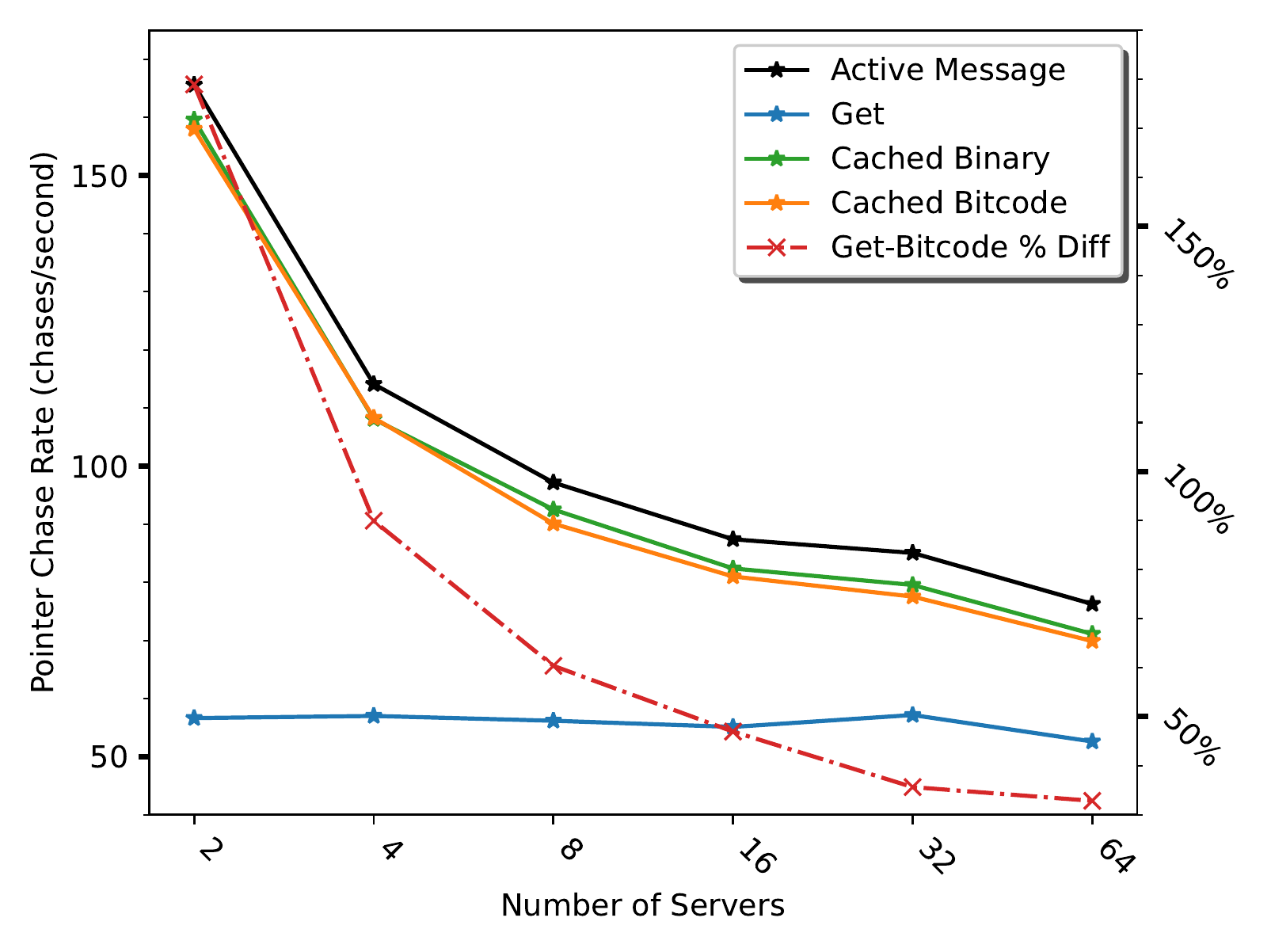}
 \centering
 \caption{\ookami{} 4096-Chase-Depth; \textbf{C/C++}: \acrfull{dapc}}
 \label{fig:fujitsu-scale}
\end{figure}

\begin{figure}[h]\includegraphics[width=0.90\linewidth]{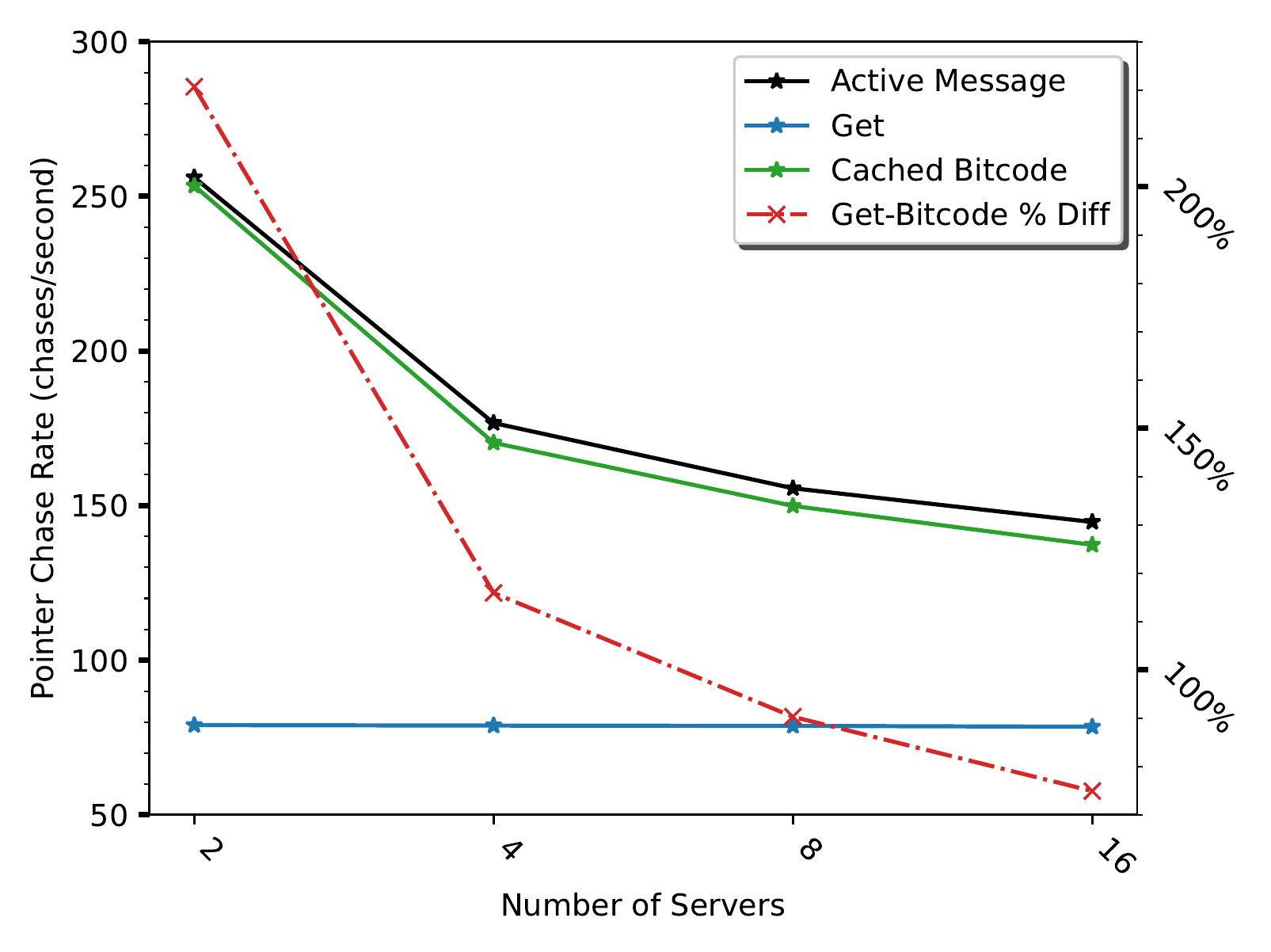}
 \centering
  \caption{\thor{} 4096-Chase-Depth; \textbf{C/C++} (Xeon Client and Servers): \acrfull{dapc}}
 \label{fig:xeon-xeon-scale}
\end{figure}

\begin{figure}[h]\includegraphics[width=0.90\linewidth]{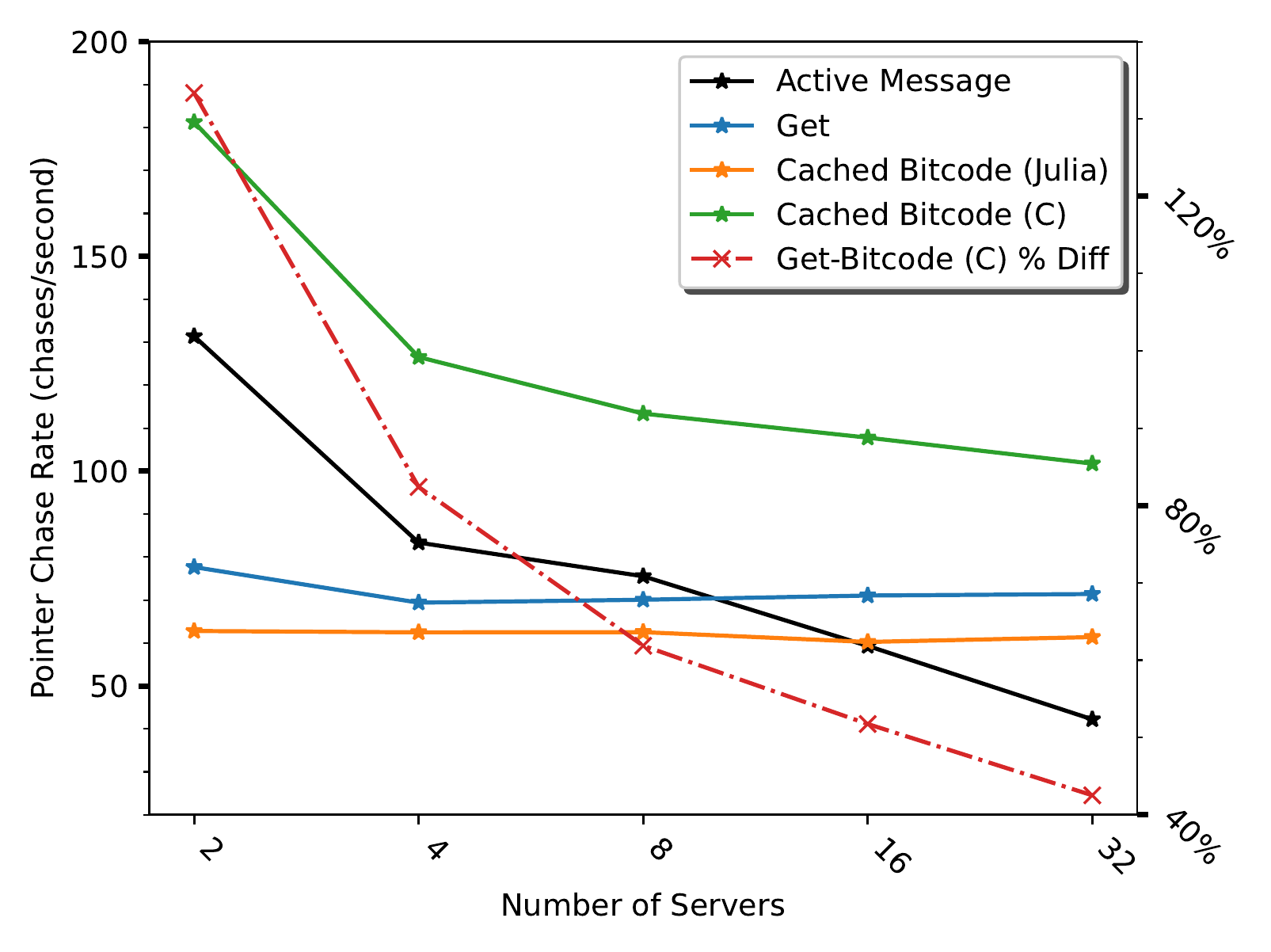}
 \centering
  \caption{\thor{} 4096-Chase-Depth; \textbf{Julia}  (Xeon Client and \acrshort{bf2} Servers): \acrfull{dapc}}
 \label{fig:xeon-bf-scale-julia}
\end{figure}

Figures \ref{fig:xeon-bf-scale}, \ref{fig:fujitsu-scale},
\ref{fig:xeon-xeon-scale}, and \ref{fig:xeon-bf-scale-julia} show how pointer
chase rate scales with the number of servers. All three C/C++ configurations 
shown exhibit similar trends. The \acrshort{gbpc} (Get) line remains relatively flat regardless of
the number of servers. This is expected because each lookup performs a get
operation that returns to the client and the number of lookups is tied to the
depth. For \am{} and \ifunc{} bitcode, the code only issues a network operation
when the value of the next lookup resides in a different server. Due to this,
the more servers the system has, the more network operations will occur. These
network operations are slower than the recursive call used when the next lookup
resides locally, leading to a lower chase rate when there are more servers. As
we saw previously, the \am{} mode performs between 3\% and 7\% better than the 
\ifunc{} Bitcode mode.

The Julia configuration (\autoref{fig:xeon-bf-scale-julia}) has similar behavior
to the C/C++ configurations with some differences worth
highlighting. In particular the performance of bitcode generated from Julia is
surprisingly constant across the scaling experiment and warrants further
investigation. In addition, Julia driving the bitcode generated from C is demonstrating
excellent performance.

%% file: content/conclusions.tex
\section{Conclusions}
In this work we described design of the \ccc{} framework and demonstrated a
performance evaluation using microbenchmarks and the \acrshort{xrdma} pointer
chase application. The framework provides the infrastructure for moving code in
binary or LLVM bitcode representation. The framework was evaluated on the
\ookami{} Fujitsu system and the \thor{} system with Intel CPUs and
\acrlong{bf2} DPUs. We showed that our framework outperforms a classical
\acrshort{rdma}-GET pointer chase implementation. We confirmed that the overhead
of the framework is comparable to that of \am{} semantics while providing a
greater level of flexibility and programmability. We also demonstrated the
framework's integration with C/C++ and Julia-based applications. While the
performance for our Julia implementation is not yet optimal, this highlights a
couple of possibilities. Firstly, code written in Julia can make use of
high-performance \ifuncs{} generated from C to allow for dynamic code execution
with high performance. Secondly, Julia as an \ifunc{} will require a
deeper investigation into the performance issues, but can enable the usage of
high-level language features, Julia's code generation capabilities as well as
use of the rich libraries available in Julia. In particular, we could conceive
the usage of machine-learning and online-statistics libraries written in Julia
for data processing on DPUs.

The \ccc{} framework and all the benchmarks used for this paper are open source and available on GitHub \cite{three-chains-github}.

%% file: content/ack.tex
\section{Acknowledgments}
The authors would like to thank the LANL for their continued support of this
project. In addition, we would like thank Jon Hermes, Alexandre Ferreira, and
Eric Van Hensbergen for their review of the paper and code.

We thank Gilad Shainer, David Cho, and HPC Advisory Council for providing access
to Thor system.  The authors would also thank Stony Brook Research Computing and
Cyberinfrastructure, and the Institute for Advanced Computational Science at
Stony Brook University for access to the Ookami computing system, which was made
possible by a \$5M NSF grant (\#1927880).
We gratefully acknowledge funding from NSF (grants OAC-1835443, OAC-2103804, AGS-1835860, and AGS-1835881), DARPA under agreement number HR0011-20-9-0016 (PaPPa). This research was made possible by the generosity of Eric and Wendy Schmidt by recommendation of the Schmidt Futures program, by the Paul G. Allen Family Foundation, Charles Trimble, Audi Environmental Foundation. This material is based upon work supported by the Department of Energy, National Nuclear Security Administration under Award Number DE-NA0003965. The views and opinions of authors expressed herein do not necessarily state or reflect those of the United States Government or any agency thereof. The U.S. Government is authorized to reproduce and distribute reprints for Government purposes notwithstanding any copyright notation herein.